\documentclass[prb,twocolumn,floatfix,amssymb]{revtex4}
\usepackage{times}
\usepackage{graphicx}

\begin{document}

\title{Impact of Cholesterol on Voids in Phospholipid Membranes}

\author{Emma Falck}
\affiliation{Laboratory of Physics and Helsinki Institute of Physics,
Helsinki University of Technology, P.\,O. Box 1100, FI--02015 HUT, Finland}

\author{Michael Patra}
\author{Mikko Karttunen}
\affiliation{Biophysics and Statistical Mechanics Group,
Laboratory of Computational Engineering, Helsinki University
of Technology, P.\,O. Box 9203, FI--02015 HUT, Finland}

\author{Marja T. Hyv\"onen} 
\affiliation{Wihuri Research Institute, Kalliolinnantie 4, 
FI--00140 Helsinki, Finland, and 
Laboratory of Physics and Helsinki Institute of Physics, 
Helsinki University of Technology, P.\,O. Box 1100, FI--02015 HUT, Finland} 

\author{Ilpo Vattulainen}
\affiliation{Laboratory of Physics~and~Helsinki Institute of Physics,
Helsinki University of Technology, P.\,O. Box 1100, FI--02015 HUT, Finland}

\date{\today}

\begin{abstract}
Free volume pockets or voids are important to many biological 
processes in cell membranes. Free volume fluctuations are a 
prerequisite for diffusion of lipids and other macromolecules in 
lipid bilayers. Permeation of small solutes across a membrane, 
as well as diffusion 
of solutes in the membrane interior are further examples of 
phenomena where voids and their properties play a central role.
Cholesterol has been suggested to change the structure and 
function of membranes by altering their free volume properties. 
We study the effect of cholesterol on the
properties of voids in dipalmitoylphosphatidylcholine (DPPC)
bilayers by means of atomistic molecular dynamics simulations.
We find that an increasing cholesterol concentration reduces the total amount
of free volume in a bilayer. The effect of cholesterol on individual 
voids is most prominent in the region where the steroid ring structures
of cholesterol molecules are located. Here a growing cholesterol content
reduces the number of voids, completely removing voids of the size
of a cholesterol molecule. The voids also become more elongated. 
The broad orientational distribution of voids observed in pure DPPC
is, with a 30\,\% molar concentration of cholesterol,
replaced by a distribution where orientation along 
the bilayer normal is favored. Our results suggest that instead 
of being uniformly distributed to the whole bilayer, these effects are 
localized to the close vicinity of cholesterol molecules.
\end{abstract}

\maketitle

\section{Introduction}

In addition to space occupied by atoms of, e.\,g., 
lipid, sterol, and protein molecules, cell membranes\cite{Alb94,Blo91,Jon95} 
incorporate free space or {\it free volume}.\cite{Alm92,Bas95,Bon54,%
Gal79,Loi04,Mar96a,Raj98} Membranes may be thought of as complex 
porous structures with a varying number of free volume pockets or voids. 
The voids might assume different sizes, shapes, and orientations. 
The nature of the voids is dynamic: they can be generated or annihilated
by trans-gauche isomerizations in the hydrocarbon tails of lipids 
molecules, or less frequently, by the movement of whole
lipids.\cite{Bas95}

Voids are crucial for dynamic processes in membranes.\cite{Raj98}
Local free volume fluctuations in bilayers are supposed to give rise
to jump diffusion of the lipids that constitute the membrane 
bilayer.\cite{Alm92,Gal79} Another important process where voids
play a central role is the diffusion of small solutes within or 
across membranes. For instance the electron transport in mitochondria
and chloroplasts is primarily influenced by the availability
of voids for the diffusion of the electron carrying quinone 
molecule.\cite{Mat93} Passive transport of water, oxygen, small organic 
molecules, and small ions to and from the cell across the plasma 
membrane is an important process involving 
voids.\cite{Bas95,Bem04,Jed03b,Mar94,Mar96b,Raj98,Sto96} Further, 
general anesthesia might be partially explained by changes
in the packing and void distribution of membranes caused by the 
partitioning and diffusion of anesthetics in membranes. 
These changes have been suggested to lead to modifications in the 
lateral pressure profile of the membrane, and will consequently alter 
the structure and function of the membrane proteins.\cite{Can99,Eck01} 
Voids are also important for the interpretation of fluorescence
anisotropy measurements.  Decreased fluorescence anisotropy caused by 
the increased mobility of fluorescent probes such as diphenylhexatriene 
(DPH) or pyrene could be due to generation of voids.\cite{Raj98,Rep04}
Finally, voids are believed to be involved in 
thinning transitions in smectic membranes.\cite{Jeu03}

There have been several attempts to formally describe the 
effect of voids on the dynamic processes in cell membranes 
or the simpler one- and multi-component lipid bilayers.
The first free volume and free area theories appeared 
decades ago.\cite{Coh59,Gal79,Alm92} Free area theories 
in particular have been used for interpreting the 
mobilities of lipid or sterol molecules in model 
bilayers.\cite{Alm92,Gal79} The basis of these 
theories is that a particle attempting a jump 
needs, in addition to a sufficient activation energy, a large 
enough empty site to jump to. The diffusion coefficient is therefore
thought to depend on the ratio of the close-packed
area of the diffusing particle and the average free area per 
particle.\cite{Alm92}

Free area theories have certain shortcomings. They are two-dimensional, 
i.\,e., it is assumed that the free volume properties do not vary
significantly in the direction of the bilayer normal. Unfortunately,
bilayers are remarkably heterogeneous in the normal direction,
and the amount of free volume varies strongly across a 
bilayer.\cite{Fal04,Mar96a} More important, free area 
theories are mean-field, that is, all predictions are based on 
the average free area available per lipid. It is 
impossible to distinguish between very few large voids and numerous tiny 
ones. Such a distinction should be highly relevant from the point of 
view of, e.\,g., solute diffusion; unhindered motion 
within a substantial void is expected to differ from jumps 
between isolated voids. Hence, free area theories
should not be used for quantitative predictions. Of late, there 
have been a few attempts to develop
more sophisticated free volume theories.\cite{Mit99,Xia99}
These analytical approaches have enhanced our understanding 
of the impact of voids on dynamic processes.

A detailed understanding of the microscopic properties of voids and 
their role in dynamical processes will, however, not emerge from 
analytical considerations, but can be acquired using computer 
simulations. In two early molecular dynamics (MD) studies, Bassolino-Klimas 
et al.\cite{Bas93,Bas95} focused on the diffusion 
mechanisms of benzene molecules in phospholipid
bilayers at different temperatures in the fluid phase.
The size distribution of voids was used to explain the
motion of the benzene molecules. Further, the size
distribution was found to depend on temperature.
Marrink et al.\cite{Mar96a} performed a detailed analysis
of the properties of voids in a dipalmitoylphosphatidylcholine
(DPPC) bilayer in the fluid phase. The size, shape, and orientation 
of the voids were found to vary significantly in the direction of the 
bilayer normal. More recently, Shinoda et al.\cite{Shi04} considered 
the effect of chain branching on the permeability of lipid bilayers,
and briefly addressed the issue of voids. They concluded that chain 
branching would seem to reduce the probability of observing large voids.
Jedlovszky et al.\cite{Jed03b} have briefly considered the 
impact of cholesterol on the number of voids in dimyristoylphosphatidylcholine 
(DMPC) bilayers. In all, computer simulations can be considered a useful tool
for studying voids.

In this computational study, we focus on the impact of cholesterol
on the properties of voids in lipid bilayers. Cholesterols
are rigid molecules that seem to enhance the properties of 
cell membranes and allow for wider variations in the lipid
composition of the membrane.\cite{Vis90} Important effects
of a finite cholesterol concentration on bilayers in the fluid state
include changes in passive permeability of small 
solutes\cite{Jed03b,Xia99} and suppressed lateral diffusion 
of phospholipids in bilayers containing cholesterol.\cite{%
Alm92,Fal04,Gal79,Hof03,Pol01,Vat03} These effects suggest that
cholesterol occupies free volume in lipid bilayers.\cite{Alm92,Fal04,Gal79}
The void-filling capacity of cholesterol is also thought to be
relevant for the formation of membrane domains enriched
in glycosphingolipids and cholesterol.\cite{Pra00} Our intention is
to study in microscopic detail how the presence of cholesterol molecules
influences the properties of voids in phospholipid bilayers.

We have performed 100\,ns MD simulations of DPPC\,/\,cholesterol 
bilayers in the fluid state, with cholesterol concentrations
ranging from 0 to 29.7\,mol\,\%.\cite{Fal04} The first step in
characterizing the voids is to single out the free volume by mapping 
the bilayer on a rectangular three-dimensional grid and checking 
which grid points lie outside the van der Waals radii of any 
atoms in the system. The individual free volume pockets 
or voids are identified by using a weighted tree-based union\,/\,find 
algorithm with path compression adapted from 
Ref.~\onlinecite{New01}. Finally, the properties of the 
voids are analyzed with the aid of principal component analysis,\cite{Sha96} 
a technique that yields information on the shape and 
orientation of voids. Combining these methods, we 
investigate the effects of cholesterol, permeant size, and location 
in the bilayer on the properties of voids. We first
focus on the size distribution of voids, considering also
percolation of free volume. The shape and
orientation of voids, and the possible size
dependence of these properties, are also examined. Finally, we
look into local effects on voids in the close vicinity of 
cholesterol molecules.

These studies are qualitative rather than quantitative in nature.
Our intention is to identify trends and describe new phenomena, 
simultaneously attempting to shed light on the microscopic 
reasons behind them.

Our results suggest that replacing phospholipid molecules by
cholesterols reduces the total volume of the bilayer. In addition,
both the total free volume and the free volume fraction decrease
with an increased cholesterol content. As for the void distribution, 
the effects of cholesterol on the properties of voids 
are most prominent in the region where the rigid 
steroid ring parts of cholesterol molecules reside. Other parts of the bilayer 
are affected to a lesser degree. In the region with
the ring structures, the presence of cholesterol 
reduces the number of voids, completely removing
voids of its own size. The voids become more elongated
and instead of a broad orientational distribution, 
which is observed in pure phospholipid bilayers, with 29.7\,\% cholesterol
voids are preferentially oriented along the bilayer normal. We also
find that the effect of cholesterol is local, i.\,e., the effects are most
apparent in the close vicinity of cholesterol molecules.

\section{Model and Simulation Details}

The model systems we studied are fully hydrated lipid bilayers 
consisting of 128 macromolecules, i.\,e., DPPCs and cholesterols, 
and 3655 water molecules. In this contribution we focus on 
a pure DPPC bilayer and DPPC\,/\,cholesterol bilayers 
with five different cholesterol molar fractions: 
$\chi = 0$\,\%, 4.7\,\%, 12.5\,\%, 20.3\,\%, and 29.7\,\%.
The underlying MD simulations have been described and validated 
elsewhere,\cite{Fal04} and in the following we shall 
merely summarize the most important details. The previously published
results have been shown to be in good agreement with 
experimental studies.\cite{Fal04}

The united atom force fields for the DPPC and cholesterol molecules 
were adopted from earlier studies.\cite{Ber97,Hol01,Tie96} 
As an initial configuration for the pure DPPC bilayer we used the
final structure of run E in Ref.~\onlinecite{Tie96}. 
For systems with finite cholesterol concentrations, the
initial configurations were constructed by replacing randomly selected
DPPC molecules with cholesterols such that same number of DPPC molecules
was replaced in each monolayer. To fill the free volume
left by replacing DPPC molecules by the somewhat smaller cholesterols, 
the system was thoroughly equilibrated, see Ref.~\onlinecite{Fal04} 
for details. The bilayer was aligned such that it lies in the 
xy plane. The bilayer normal is thus parallel to the z-axis.

The MD simulations were run at a temperature $T = 323$\,K
using the GROMACS\cite{Lin01} molecular simulation software. 
After the initial equilibration, the system was kept in the
$NpT$ ensemble at a pressure of 1\,bar using the 
Berendsen barostat and thermostat.\cite{Ber84} The duration of 
each simulation was 100\,ns and the time step 
was chosen to be 2.0\,fs. For long-range electrostatic 
interactions we used the Particle-Mesh Ewald (PME) 
method,\cite{Ess95a} which has been shown to yield reliable 
results for bilayers.\cite{Pat03a,Pat03b}

\section{Methods}

The details of the technique outlined in this section, along with 
a computer code, will be published elsewhere.\cite{Fal05}

\subsection{Finding Free Volume: Grid Approach\label{sect_grid}}

To locate the free volume, the bilayers were mapped onto three-dimensional
rectangular grids, an approach outlined in Ref.~\onlinecite{Mar96a}. 
The grid elements outside the van der Waals radii of 
any atomic groups were designated as free volume and those inside as
occupied volume. The van der Waals radius of a given 
atomic group was taken to be half of the 
shortest distance where the Lennard--Jones interaction of that 
atomic group with an identical atomic group is zero. 
The Lennard--Jones parameters were extracted from the 
specification of the force field.

The grid thus constructed stretched, in the normal direction,
between the points where the density of water starts to deviate
from its bulk value. The number of grid elements in each direction
was chosen such that the linear size of an element was 0.075\,nm 
on the average. As the area of the bilayer decreases and the
thickness increases with an increasing cholesterol content,\cite{Fal04} 
the number of elements had to be set separately for each system 
with a different cholesterol concentration. For instance in the case
of pure DPPC, the total area of the bilayer is 41.9\,nm$^2$ and the distance
between the points where the density of water
no longer corresponds to the bulk value is 5.8\,nm. 
With 29.7\,\% cholesterol, the corresponding values are 26.9\,nm$^2$
and 6.4\,nm. In the case of pure DPPC, the number of grid elements in the
x and y directions was taken to be 86. The number of elements in the
normal or z direction was 78. The corresponding values
for the bilayer with 29.7\,\% cholesterol were 74 and 82.

We also used finer grids to make sure the results were not 
influenced by the resolution. In turns out that when
the linear size of a grid element is reduced to 0.05\,nm, all
quantities we have computed change only slightly. The 
changes are clearly within the error bars originating
from a finite simulation time and correlated samples.

In constructing the grids, in the spirit of Ref.~\onlinecite{Mar96a},
we took into account the finite size of small
solutes diffusing in bilayers. Due to the finite size, not 
all grid elements outside the van der Waals radii
of the atoms can be accessed by the center of mass (CM) of the
diffusing solute. Therefore, in addition to the 
empty free volume, i.\,e., all the free volume outside the van der 
Waals radii of the atoms, we also studied the so-called accessible 
free volume. Accessible free volume is calculated by adding the van der Waals
radius of the diffusing solute to the van der Waals radii of the
atoms in the bilayer, and corresponds to the free volume which is accessible
to the CM of the diffusing solute. Several solute sizes
with radii $r$ ranging between 0 and 0.2\,nm were studied. For comparison,
the effective van der Waals radii of bare sodium, chloride, and 
potassium ions, as well as those of water and oxygen molecules, 
are between 0.1 and 0.2\,nm. The same applies for example to the 
general anesthetic Xenon.

The resulting grids were used to compute the average free volume fraction
as a function of the distance from the bilayer center. From these
it was easy to extract the total free volume in each system. 
The grids were also the starting point for studying the properties of
voids for each system.

\subsection{Identifying and Characterizing Voids}

To characterize the distribution of voids, we need to identify all voids,
that is, all clusters of free volume, in our grid. Computing and 
characterizing the distribution of contiguous clusters
of occupied or unoccupied sites is a well-known problem,
usually solved by a union\,/\,find approach.\cite{Cor01,New01}
It works as follows. The sites, in this case the empty grid elements, 
must be traversed. Whenever we arrive at a new site, 
a find operation is  performed, i.\,e., we check which 
clusters the nearest neighbor sites, if unoccupied, 
belong to. The union operation follows: if there are unoccupied nearest 
neighbor sites that belong to different clusters, those clusters 
are amalgamated. Otherwise, no union of two clusters is needed.

Modern union\,/\,find algorithms are weighted, tree-based, and
use path compression.\cite{Cor01,New01} Each cluster is stored
as a tree such that the nodes of the tree are the sites
of the cluster. Every cluster has a root that corresponds to the
root node of the tree, and the other nodes have pointers to either the root
or to another node in the same tree. By following a succession of 
such pointers we can locate the root of any tree. The find operation
therefore consists of traversing trees by following pointers to 
the root nodes. If the nearest neighbor sites of a new site
lead to the same root node, the sites belong to the same cluster.
The find operation can be implemented such that after 
the traversal is completed, the pointers along the path all
point directly to the root node. This is called path-compression, 
and it speeds up future traversals. The union of two clusters is achieved
by adding a pointer from the root node of one tree to the root
of the other. Hence, one of the trees becomes a subtree of the other.
This is done in weighted fashion such that the smaller tree becomes
a subtree of the larger tree, which is easily accomplished by storing the
size of each cluster in the corresponding root node.

Even though the above algorithm may seem a little complicated,
it can be implemented, using recursion, in a very elegant and 
concise manner.\cite{New01} Moreover, its performance is in most 
cases far superior\cite{New01} to simple relabeling algorithms such as 
that introduced in Ref.~\onlinecite{Hos76}. We chose therefore
to adapt the weighted tree-based union-find code with path
compression given in Ref.~\onlinecite{New01}.

The end result of the union\,/\,find algorithm for a given configuration
is a tree structure containing the information on how the unoccupied
sites with known coordinates are connected to each other. From 
the tree structure we constructed arrays where 
each array element corresponds to a free volume cluster or void
and contains the relevant properties of the void. Such properties
are the volume of the void, the location of its CM, and
its maximum and minimum coordinates in each direction. Further,
we stored the covariance matrix extracted from the coordinates of the
grid elements that belong to the void. The covariance matrix
can be used to characterize the shape and orientation of the void,
as described in the next section.

\subsection{Principal Component Analysis\label{sect_pca}}

The shape and orientation of the voids were studied using principal 
component analysis (PCA).\cite{Sha96} PCA is a versatile statistical method,
which is widely used in analysis and compression of multivariate data, 
signal processing, and neural computing. Here we 
shall confine ourselves to the case of three-dimensional data.

In applying PCA, we view each void as a three-dimensional ellipsoidal
cloud. The principal component vectors are orthogonal and 
represent the directions with maximum variability. In other words,
the principal components define the axes of the ellipsoidal void
such that the first principal component is the longest axis, the second
is the next longest axis, and the third is the shortest axis.

The principal components may be computed as follows.
For a void consisting of $N$ grid elements with coordinates 
$\{\mathbf{r}_i \}_{i=1}^N$, we first compute the mean (or CM) 
$\mathbf{R}$ and the covariance matrix $\mathbf{C}$. The
diagonal of the covariance matrix contains the variances of the 
Cartesian components of $\mathbf{r}_i$. The off-diagonal 
matrix elements represent correlations between 
the different components of $\mathbf{r}_i$. 
The covariance matrix is symmetric, and by 
finding its eigenvalues and eigenvectors we 
find an orthogonal set of basis vectors. By ordering the eigenvalues, 
we obtain an orthogonal basis with the first eigenvector 
having the direction of the largest variance of the data and so on. This is 
the set of uncorrelated principal components, and simultaneously,
the axes of our ellipsoidal void. The square roots of the
eigenvalues, in turn, are proportional to the lengths of the axes.

The shape of a void can be deduced from the eigenvalues $\sigma_i^2$ of the
covariance matrix. We chose to focus on the ratios of the square roots of 
the eigenvalues, denoted by $\sigma_i / \sigma_j$. In case  
$\sigma_1 / \sigma_2 \approx \sigma_2 / \sigma_3 \approx 1$, the void
is spherical, otherwise it is elongated. The principal components 
were used to calculate the orientation of each elongated or 
non-spherical void. We chose to regard
all voids with $\sigma_1 / \sigma_3 < 1.5$ as spherical. Even though
this is not a very strict definition of sphericity,
most voids appeared to be non-spherical. Further, this definition
was found to be robust, since changing the somewhat arbitrary 1.5
to lower values had virtually no effect on the number of spherical 
clusters. For the elongated voids, we monitored the angle $\phi$ 
between the longest axis and the bilayer normal.

As for the practical details of the PCA calculation, we computed 
the eigenvectors and eigenvalues using singular value 
decomposition. Further, we only used PCA for voids that 
were larger than $4 \times 10^{-3}$\,nm$^3$
and smaller than $0.13$\,nm$^3$. The very small voids were 
rejected, because it is somewhat artificial to discuss the shape
of an object consisting of one or two rectangular grid elements.
The large voids were not considered, because their shapes
were, in most cases, not ellipsoidal, but rather more complicated with
branches and compartments. Applying PCA to these voids would have 
been meaningless.

\subsection{Four Region Model\label{sect_four_regions}}

The bilayer is highly heterogeneous in the normal direction, and 
both the free volume fraction and the properties of the voids are
anticipated to vary with the position along the bilayer normal.\cite{Mar96a}
It is therefore not sensible to study the void distribution
averaged over the whole bilayer, but to discretize the bilayer
into regions with more homogeneous compositions. The
properties of voids can then be studied in these separate regions.
We chose to analyze the data in terms of the four region model 
introduced by Marrink et al.\cite{Mar94,Mar96a}

The original partition into four regions was slightly modified 
to better probe the effects of cholesterol
on the properties of voids, the main modification being
that Region 3 was taken to be the part of the bilayer where
the density of cholesterol steroid ring structures is high.
We expect the effects of cholesterol on the voids 
to be most prominent in this region, as the steroid rings
rigidify the disordered tails of the DPPC molecules,  
and enhance the molecular packing.\cite{Fal04,Hof03}

\begin{figure}[h]
\centering
\includegraphics[width=\columnwidth,clip=true]%
{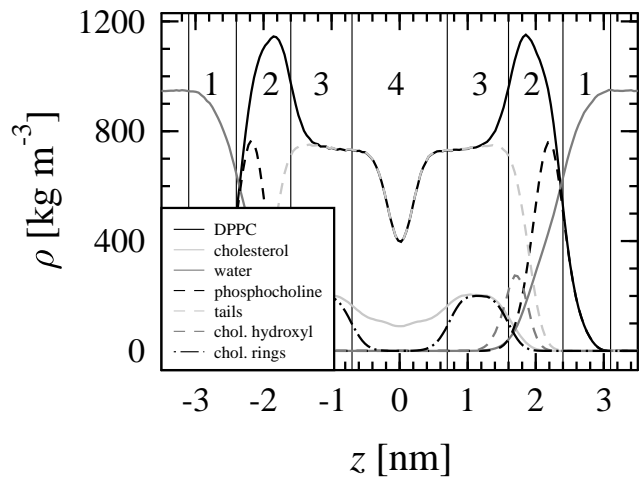}
\caption{\label{fig_regions}Mass density profiles of molecules and 
atomic groups at $\chi = 20.3$\,\%. The mass density of the cholesterol 
hydroxyl groups has been scaled by a factor of ten. 
The division into four regions,
indicated here by numbers $1-4$, is based on the mass density 
profiles, see Sect.~\ref{sect_four_regions}.}
\end{figure}

Figure~\ref{fig_regions} illustrates how we have divided the 
bilayer into four distinct regions. Region 1 
ranges from the point where the density of water starts to 
deviate from the bulk value to the point where the densities of DPPC and 
water are equal. This region therefore contains mostly water
molecules and a low density of DPPC headgroups. Region 2
extends from the point where the densities of water and DPPC
coincide to the point where the density of cholesterol ring
structures is half of its maximum plateau value.\cite{Fal04}
This region contains mainly parts of DPPC molecules: the choline,
phosphate, and glycerol densities peak in this region, and
there is also a finite density of DPPC tails. Further, the hydroxyl groups
of cholesterol reside mainly in this region. Region 3 is defined
between the points where the cholesterol steroid ring density
assumes half of its maximum value, and contains in addition  
the middle parts of the DPPC tails. The remaining
part of the bilayer is Region 4, i.\,e., this region is in the center of
the bilayer. Here we expect to find a low density of DPPC and 
cholesterol tails, with the end methyl groups being mainly
located in this region.

The boundaries between the regions vary with cholesterol concentration.
This is due to the fact that an increasing cholesterol content
changes the densities of the various components such that
they are shifted away from the bilayer center.\cite{Fal04} Hence, in the
case of pure DPPC, the boundaries from the outer boundary of Region 1 to the
boundary between Regions 3 and 4 were set to 2.9\,nm, 2.1\,nm, 1.4\,nm, 
and 0.6\,nm from the bilayer center. With 29.7\,\% cholesterol, 
the corresponding values were 3.2\,nm, 2.5\,nm, 1.7\,nm, and 0.8\,nm.

\subsection{On Percolation Theory\label{sect_perc}}

Percolation theory can be used to predict certain
properties of clusters of occupied or unoccupied sites or bonds 
in a lattice.\cite{Sta92} The applications of percolation theory range
from modeling forest fires or the distribution 
of oil and natural gas in the porous rock in oil reservoirs to
studying, say, voids in lipid bilayers. The properties of the
clusters are studied as a function of the fraction of occupied or 
unoccupied sites or bonds, depending on the application. In our
case, we are interested in the free volume fraction
\begin{equation} \label{eq_free_vol_frac}
p \equiv \frac{\langle V_\mathrm{free} \rangle}
{\langle V_\mathrm{tot} \rangle},
\end{equation}
where $\langle V_\mathrm{free} \rangle$ is the average free volume in a bilayer
and $\langle V_\mathrm{tot} \rangle$ is the total volume of the bilayer. 
An important free volume fraction is the percolation threshold $p_\mathrm{c}$,
at which there is a free volume cluster stretching in the 
x, y, or z direction from one side of the system to the opposite side. The size
distribution of clusters depends heavily on $p$ and the vicinity of the 
percolation threshold. At $p \geq p_c$, there is a free volume 
cluster that spans the whole
bilayer, and a large fraction of all empty grid elements belong to this 
percolating cluster. When $p < p_c$, there is no such percolating cluster.

The detailed form of the size distribution of clusters cannot, 
in most cases, be predicted analytically.\cite{Sta92} A general 
scaling form for the size distribution has been postulated, 
and the non-universal details such as the form of the scaling 
function can be extracted from computer simulations. In certain
limiting cases, however, the exact form of the scaling function 
becomes irrelevant.\cite{Sta92}

Very close to the percolation threshold $p_\mathrm{c}$ or for very small 
clusters, the size distribution $N(V)$ scales with the cluster volume 
as\cite{Sta92}
\begin{equation} \label{eq_perc_at_pc}
N(V) \sim V^{-\tau}. 
\end{equation}
Here $\tau$ is a universal scaling exponent, which in three dimensions
assumes the value 2.18.\cite{Sta92} This scaling law is valid for
clusters whose linear size is smaller than the correlation length $\xi$,
which can be thought of as proportional to a typical
cluster diameter. At the percolation threshold $\xi$ approaches infinity,
and Eq.~(\ref{eq_perc_at_pc}) is valid for all cluster sizes.

Away from $p_\mathrm{c}$ and for large clusters the size distribution 
scales as follows:\cite{Sta92}
\begin{equation} \label{eq_perc_away_from_pc}
N(V) \sim \left\{ \begin{array}{ll} 
V^{-\theta} \exp(-c'' V) & \mbox{$p < p_\mathrm{c}$}, \\
V^{-\theta'} \exp(-c''' V^{1-1/d}) & \mbox{$p > p_\mathrm{c}$}.
\end{array}
\right. 
\end{equation}
Here $d$ is the dimensionality of the system, $c''$ and 
$c'''$ are proportional to $p-p_\mathrm{c}$, 
and $\theta$ and $\theta'$ are universal exponents. In three 
dimensions $\theta = 3/2$ and $\theta' = -1/9$.\cite{Sta92} 
This scaling form applies for clusters whose 
linear size is larger than $\xi$, in practice
for almost all clusters in systems that are not in the near vicinity of 
$p_\mathrm{c}$. Because of the $p-p_\mathrm{c}$ dependence of $c''$ and 
$c'''$, further away from $p_\mathrm{c}$
the crossover from power-law to exponential behavior is 
observed at smaller values of $V$.

We have computed size distributions for systems with different
cholesterol concentrations and penetrant sizes. In interpreting 
the size distribution data, it is useful to know
how close to the percolation threshold a given system is.
As opposed to infinite systems, in a finite system where the 
linear dimension in the z direction cannot be increased, 
only apparent percolation thresholds can be calculated.\cite{Mar96a}
Computing apparent percolation thresholds is not a problem, since
this is the threshold a permeant solute will experience.\cite{Mar96a}
The apparent percolation threshold can be defined as the free volume fraction
at which at least 50\,\% of all systems percolate.\cite{Mar96a} 
Alternatively, $p_\mathrm{c}$ can be extracted from maxima in 
the second moment of the cluster size distribution, 
computed over all clusters but the percolating one.\cite{Sta92}

\begin{figure}[h]
\centering
\includegraphics[width=\columnwidth,clip=true]%
{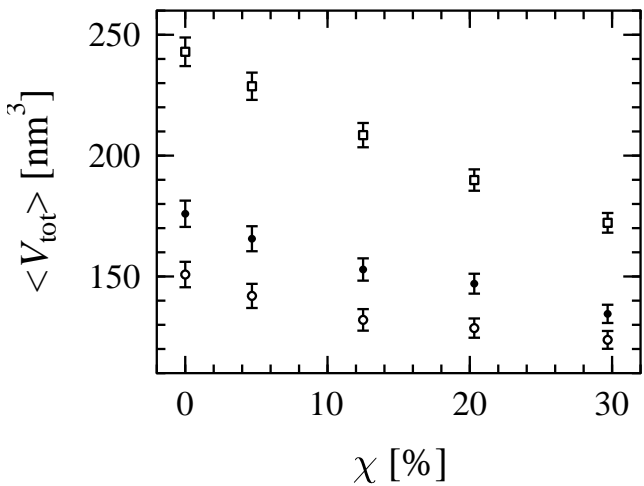}
\caption{\label{fig_vol_tot}Bilayer volume as function of cholesterol
concentration. As there is no unique definition for the thickness of
a bilayer, the thicknesses have been measured in three ways: 
between the points where the mass density of water starts to deviate 
from the bulk value ($\square$), between the points where
the mass density of phospholipids and water are equal ($\circ$), 
and between the maxima in the total electron densities ($\bullet$).}
\end{figure}

Unfortunately, in bilayer systems we cannot automatically expect 
Eqs.~(\ref{eq_perc_at_pc}) and (\ref{eq_perc_away_from_pc})
with three-dimensional critical exponents
to describe the behavior of the size distribution at and
off the apparent percolation threshold. Predicting the form of the 
size distribution is in fact much more difficult for bilayer systems than 
for isotropic bulk systems in two or three dimensions. 
First, the bilayer is heterogeneous in the normal direction,
and consequently, the free volume fraction varies with the
position along the bilayer normal, i.\,e., $p = p(z)$, see 
Fig.~\ref{fig_scaled_free_vol_prof}. The bilayer center or Region 4, with
the highest free volume fraction in the whole bilayer,\cite{Fal04,Mar96a} 
could be well above the percolation threshold, while the other regions
are well below $p_\mathrm{c}$. This problem can be in part remedied by 
considering the four regions with approximately constant free volume 
fractions. Secondly, the z direction is very different from the
x and y directions, and there might well be different thresholds
for percolation in different directions. Finally, there are finite size 
effects. The dimension in the normal direction will seldom 
exceed 10\,nm in bilayer systems. This makes the system
quasi-two-dimensional rather than three-dimensional.

\section{Results}

\subsection{Total Volume and Free Volume}

It is believed that the total free volume in a lipid bilayer 
decreases when the cholesterol concentration increases.\cite{Alm92,Gal79} 
On the other hand, it is often assumed that the total volume
of the bilayer is kept approximately constant with a growing cholesterol
concentration. This is attributed to the decreasing area and 
increasing thickness of the bilayer. As a cholesterol
occupies less volume than a DPPC, the volume occupied by
phospholipids and cholesterols must diminish with an increasing 
cholesterol concentration. Water is pushed away from the bilayer
center,\cite{Fal04} and hence we expect that the volume occupied 
by water molecules should at least not grow. The constant 
total volume and a reduced volume occupied by cholesterol, 
DPPC molecules, and water would together suggest that the 
free volume should, contrary to the general belief, increase. 
To determine whether the total free volume decreases or increases, 
we have computed the total volume and total free volume as 
functions of cholesterol concentration.

The total volume is computed by multiplying the average area 
of the bilayer by the thickness of the bilayer. Because of the 
rough interface between the bilayer and the water 
phase, there is no unique definition for the thickness
of the bilayer. As measures of thickness we have used 
the distance between the points where the mass density
of water starts to deviate from the bulk density, between the points
where the mass densities of phospholipids and water coincide, and
between the maxima in the total electron densities.\cite{Fal04} 
The total volume as a function of the cholesterol concentration is shown in 
Fig.~\ref{fig_vol_tot}. It is clear that regardless of how we
define the thickness of the bilayer, the total volume {\it decreases}
with a growing cholesterol concentration.

Unfortunately, there is little experimental evidence 
to either confirm or contradict this result.
To our knowledge, there are no systematic studies on 
the effect of cholesterol on the phospholipid bilayer area or volume.
Monolayer studies do exist,\cite{McC03} but the applicability of 
these results to bilayers is questionable.\cite{Kaz02,Nag00}

\begin{figure}[h]
\centering
\includegraphics[width=\columnwidth,clip=true]%
{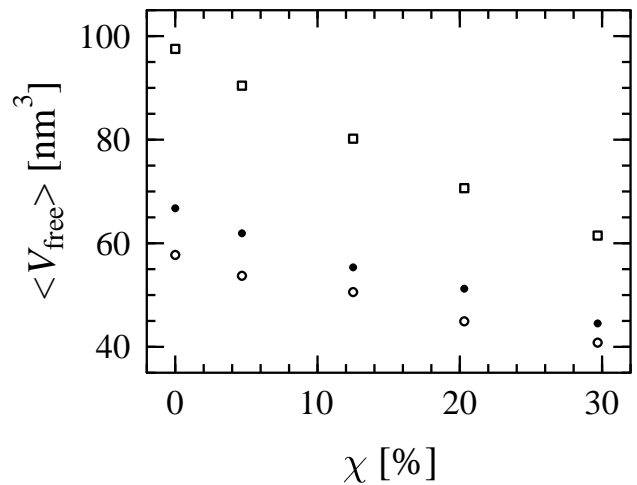}
\caption{\label{fig_vol_free}Empty free volume as function of cholesterol
concentration. The bilayer thicknesses have been defined as the distance 
between the points where the density of water starts to deviate 
from the bulk value ($\square$), the distance between the points where
the density of phospholipids and water are equal ($\circ$), 
and the distance between the maxima in the total 
electron densities ($\bullet$). The errors are of the order of a few per 
cent.}
\end{figure}

As the total volume and the occupied volume both decrease with
an increasing cholesterol concentration, it is not immediately 
clear how the free volume should be expected to 
behave. This issue can be resolved by extracting the total free 
volumes from the three-dimensional grids introduced in 
Sect.~\ref{sect_grid}. The procedure can be thought of
as integrating over the total free volumes in thin slices in the 
xy plane. The range of integration must be the thickness of the 
bilayer, defined as in the case of total volume. The results
are shown in Fig.~\ref{fig_vol_free}. It is clear that the total free 
volume decreases with an increasing cholesterol concentration,
notwithstanding how the thickness is measured. We can therefore
conclude that both the occupied and free volume decrease, together
resulting in a reduction of the total volume of the bilayer.

Experimental findings are in favor of the reduction of free volume. 
Galla et al.\cite{Gal79} have reached this conclusion 
by comparing their phospholipid diffusion 
coefficients to the predictions of free area theory. Almeida et al.\cite{Alm92}
present empirical laws for the behavior of specific volumes and thicknesses
of DMPC\,/\,cholesterol bilayers as functions of temperature. Based on these 
laws they have computed the average areas per molecule at different 
cholesterol concentrations. Combining these with close-packed areas 
for DMPC and cholesterol molecules, they have estimated the average 
free areas per phospholipid. The free areas thus obtained explain
nicely the diffusion coefficients from fluorescence recovery after 
photobleaching (FRAP) experiments.\cite{Alm92}

\subsection{Free Volume Fraction Profile}

\begin{figure}[h]
\centering
\includegraphics[width=\columnwidth,clip=true]%
{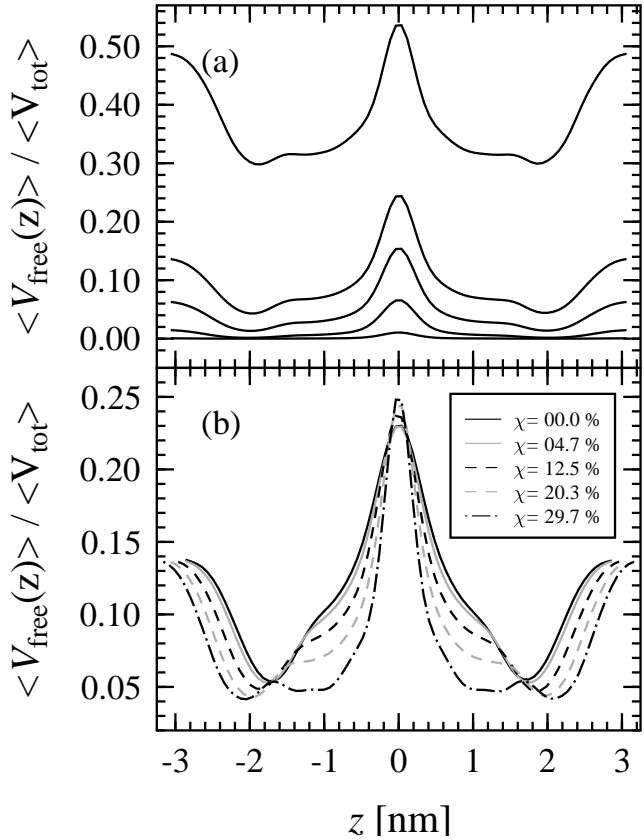}
\caption{\label{fig_scaled_free_vol_prof}Free volume fractions
as functions of position along bilayer normal. 
(a)~Empty free volume fraction and accessible free volume fractions
for bilayer with 20.3\,\% cholesterol. The penetrant radii from top
to bottom are 0.00\,nm, 0.05\,nm, 0.07\,nm, 0.10\,nm, and 0.15\,nm.
(b)~Accessible free volume fraction for different cholesterol 
concentrations for a penetrant with a radius $r = 0.05$\,nm.
In both panels the value $z = 0$ corresponds to the bilayer center, and 
the errors are of the order of a percent.}
\end{figure}

Having concluded that cholesterol reduces the total free volume,
let us now look at the behavior of free volume in different parts
of the bilayer. Free volume fraction profiles, i.\,e.,
free volume fractions as functions of the position $z$
along the bilayer normal, are displayed in Fig.~\ref{fig_scaled_free_vol_prof}.
Part (a) of the figure illustrates the effect of penetrant size
on the free volume fraction with 20.3\,\% cholesterol. 
The uppermost curve is the empty free volume fraction, while the 
bottom one corresponds to the accessible free volume fraction for
penetrants with a radius $r = 0.15$\,nm. Part (b) shows the effect
of cholesterol concentration on the accessible free volume fraction
with $r = 0.05$\,nm.

All free volume fraction profiles in Fig.~\ref{fig_scaled_free_vol_prof},
regardless of the penetrant size or cholesterol concentration, 
have maxima in the bilayer center and minima in Region 2. 
The penetrant size has a strong effect on free volume fraction: at 
least a third of the total volume at each $z$ seems to be 
empty free volume, while for penetrants with $r = 0.15$\,nm, 
apart from the bilayer center, there is virtually no accessible free volume at 
all. These findings are in good qualitative agreement with the results of
Marrink et al.,\cite{Mar96a} considering that those simulations
were performed at a temperature $T = 350$\,K and that temperature has
a significant effect on free volume.\cite{Bas95} Cholesterol 
influences the shape of the profile, reducing the free volume fractions 
in Regions 1 and 3. The decrease in Region 3 must be 
due to the presence of the rigid steroid ring structure. In 
Region 1, the origin of the effect may be the reorientation of the DPPC 
headgroups.\cite{Fal04}

\subsection{Percolation}

\begin{figure}[h]
\centering
\includegraphics[width=\columnwidth,clip=true]%
{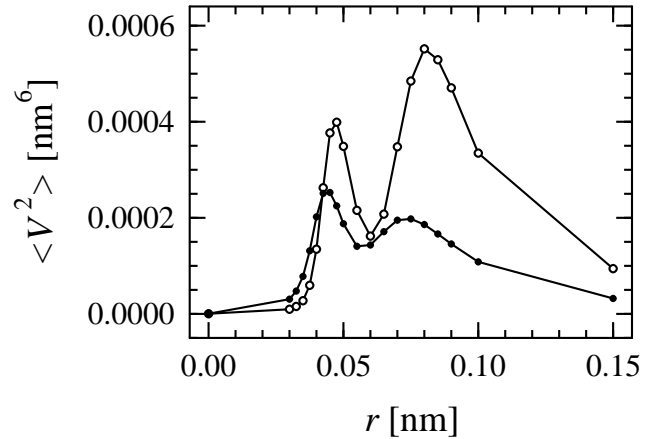}
\caption{\label{fig_secmom}Second moments of void size distributions
computed over all but percolating cluster 
as functions of penetrant radii. The open circles are for $\chi = 0$\,\%
and the filled ones for $\chi = 29.7$\,\%. The errors are of the order
of a few percent.}
\end{figure}

Examining the nature and location of the percolation transition
is essential for a proper interpretation of the size distributions.
We have located the percolation threshold by computing
the second moment of the size distribution of voids, see 
Sect.~\ref{sect_perc}. The second moments for $\chi = 0$\,\%
and 29.7\,\% as functions of penetrant radii are shown in 
Fig.~\ref{fig_secmom}. Since a maximum indicates percolation,
we can identify two percolation transitions for each cholesterol
concentration. For small $r$ percolation takes
place in x, y, and z directions. In between the two
maxima the system percolates in the plane of the bilayer, but no
longer in the normal direction. This xy percolation always
takes place in Region 4. Both of these transitions are important from a
biological point of view: the xyz percolation is relevant for
permeation across the bilayer, whereas xy percolation
in the bilayer center could be significant, e.\,g., for diffusion of quinone
within the bilayer. Finally, at large values of $r$ there is no
percolation.

The location of the transitions depends weakly on cholesterol 
concentration. The xyz percolation takes place at
$r = 0.048(2)$\,nm when $\chi = 0$\,\% and at
$r = 0.044(2)$\,nm when $\chi = 29.7$\,\%. The corresponding 
values for the xy percolation are $r = 0.080(2)$\,nm and 
$r = 0.073(2)$\,nm. A larger fraction of cholesterol
therefore seems to imply that slightly smaller penetrants 
perceive the system as percolating.

As for percolation thresholds, it is meaningless to define 
anything such for the xyz percolation, since the free 
volume fraction varies strongly along the bilayer normal. 
The xy percolation in the more homogeneous Region 4 is quite 
another matter. In this case we can,
using plots similar to Fig.~\ref{fig_scaled_free_vol_prof}, map
the penetrant sizes for which percolation takes place to free 
volume fractions. The threshold for xy percolation
is $p = 0.06$ at all cholesterol concentrations. Cholesterol
does not alter the percolation threshold itself, but it changes slightly
the free volume fraction for a given penetrant size. This causes
the weak $\chi$ dependence of the location of the xy percolation
seen in Fig.~\ref{fig_secmom}.
The value of $0.06$ is in good agreement with the corresponding
threshold calculated by Marrink et al.\, for pure DPPC
and thus equals their value for soft polymer as well.\cite{Mar96a}

\subsection{Size Distributions\label{sect_sizedistr}}

The size distributions are unnormalized to better illustrate
the behavior of the absolute number of voids in a bilayer
of a given composition. Therefore $N(V)$ is the average number
of voids of volume $V$. As the four regions have very distinct
material properties,\cite{Mar94,Mar96a,Mar96b} 
we have considered the size distributions
separately in each region. Further, the vicinity of the
percolation transitions is significant for the form
of the distribution. We have identified two cases
that represent the most biologically relevant regimes:
$r = 0.05$\,nm at which xy percolation takes place in Region 4
and $r = 0.09$\,nm with no percolation. The regime where
xyz percolation occurs is quite similar to the case of 
$r = 0.05$\,nm.  Besides, even though the results presented here
are qualitative rather than quantitative in nature, describing
the qualitative effect of cholesterol instead of claiming 
to establish accurate numerical values, penetrant radii smaller 
than $0.05$\,nm are unphysical.

\begin{figure}[h]
\centering
\includegraphics[width=\columnwidth,clip=true]%
{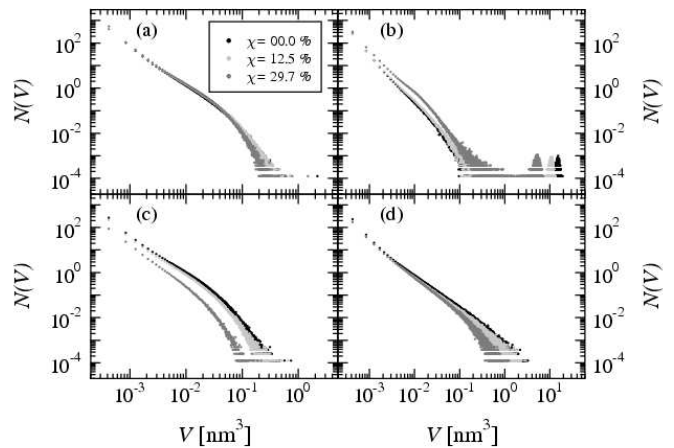}
\caption{\label{fig_num_cho}Effect of cholesterol concentration
on void size distribution.
(a)~Region 3 and $r = 0.05$\,nm.
(b)~Region 4 and $r = 0.05$\,nm.
(c)~Region 3 and $r = 0.09$\,nm.
(d)~Region 4 and $r = 0.09$\,nm.
Percolation in the plane of the bilayer is seen in panel (b)
as a separate cluster of points in the right-hand corner. 
The finite statistics lead to relative errors that 
grow with $V$. For $V < 0.01$\,nm$^3$ the errors
in $N(V)$ are smaller than a percent, and in the range
$0.01$\,nm$^3$ $< V < 0.1$\,nm$^3$ smaller than ten percent. 
If $V$ is of the order of $1$\,nm$^3$, the relative errors may 
be $100-300$\,\%. As the data are shown on a loglog scale, 
this is hardly a problem.}
\end{figure}

The effect of cholesterol on the void size distributions 
is largely restricted to those parts where the cholesterol
molecules are located. In Region 1, regardless of $r$, 
cholesterol has no impact on the size distributions.
This is not surprising, since no part of cholesterol
is present in the water phase. The effects are still very minor 
in Region 2, although the hydroxyl groups of cholesterol
are mainly located in this region. The situation in Regions 
3 and 4, where the cholesterol steroid ring structures and tails are
situated, is illustrated in Fig.~\ref{fig_num_cho}. These
are somewhat coupled, since a particularly large cluster
could extend to both regions, but is assigned to Region 4,
as its CM is located there.

The behavior of these two regions with an increasing cholesterol 
concentration depends on $r$. The case of $r = 0.05$\,nm,
representing all $r$ below the xyz percolation threshold
and above the xy threshold, is portrayed in panels (a) and (b)
of Fig.~\ref{fig_num_cho}. The most notable effect of replacing 29.7\,\% of the
DPPC molecules by cholesterols is the decreasing connectivity of 
free volume. This is manifested in the size of the percolating 
cluster, which decreases by a factor of three to $V \approx 5$\,nm$^3$, 
approximately ten times the average close-packed volume of a 
cholesterol molecule, $V_\mathrm{chol} \approx 0.459(2)$\,nm$^3$.\cite{Fal04} 
When connectivity is reduced, the number of non-percolating 
voids of all sizes in Region 4 increases: although the total free volume 
decreases, the number of voids increases approximately by a factor 
of three. The effects in Region 3 are more minor. Here
the number of voids from $V$ with radii of $0.1$\,nm to 
$V$ of the order of a tenth of $V_\mathrm{chol}$ 
increases very slightly. The number of 
large voids with $V$ close to $V_\mathrm{chol}$ decreases somewhat.

The situation is quite different when no percolating cluster is 
present, as illustrated by results computed with $r = 0.09$\,nm 
in panels (c) and (d) of Fig.~\ref{fig_num_cho}. The effects 
of cholesterol are very pronounced 
in Region 3. Comparing the case of $29.7$\,\% cholesterol to 
pure DPPC we note that there are fewer voids of all sizes.
The number of voids is reduced at least by a factor of three,
but especially for larger voids the reduction can be
as large as a factor of twenty. The largest voids with 
0.2\,nm$^3 \lesssim V \lesssim$ 0.7\,nm$^3$, 
i.\,e., voids with sizes close to $V_\mathrm{chol}$, 
are removed completely. The effects are similar, although more minor, 
in Region 4. Voids with radii up to $0.1$\,nm are unaffected,
while the number of intermediate and large voids decreases. Here
cholesterol removes the voids with 1\,nm$^3 \lesssim V \lesssim$ 3\,nm$^3$, 
larger than its own close-packed size.

Let us finally analyze Fig.~\ref{fig_num_cho} from the point of view
of percolation theory. Below the xyz percolation threshold, i.\,e.,
for both $r = 0.05$\,nm and $r = 0.09$\,nm, there is no 
percolating cluster in Region 3. If we assume that bulk 
three-dimensional percolation theory holds, in Region 3
we expect to observe power-law behavior with an exponent $\theta = 1.5$
at small and intermediate $V$ and cross-over to exponential behavior
at large $V$. This indeed seems to be the case, see panels (a) and (c).
We also note that the cross-over to exponential behavior occurs at smaller
$V$ for $r = 0.09$\,nm. This is to be anticipated, since
the larger the penetrant radius, the further away from the
xyz percolation threshold we are. In the case of 
Region 4 with $r = 0.09$\,nm, see panel (d), we are very close to the 
xy percolation threshold. At very small $V$,
power-law behavior with an exponent close to $2.1$ is observed.
The linear sizes of these small voids must hence be smaller than the
correlation length $\xi$. At intermediate and large $V$, the 
power-law behavior can still be observed, with an exponent close 
to $1.5$. Pure DPPC, which is closer to the percolation 
threshold than bilayers with finite cholesterol concentrations, 
shows no cross-over to exponential behavior, while with 
finite $\chi$ we observe exponential behavior for the largest $V$.
The behavior in Region 4 above the xy percolation
threshold, see panel (b), is difficult to interpret, as we are in between two
percolation thresholds. At small and intermediate 
$V$ we again observe power-law behavior with an exponent close 
to $2.2$. At larger values of $V$, the situation is less clear.

\begin{figure}[h]
\centering
\includegraphics[width=\columnwidth,clip=true]%
{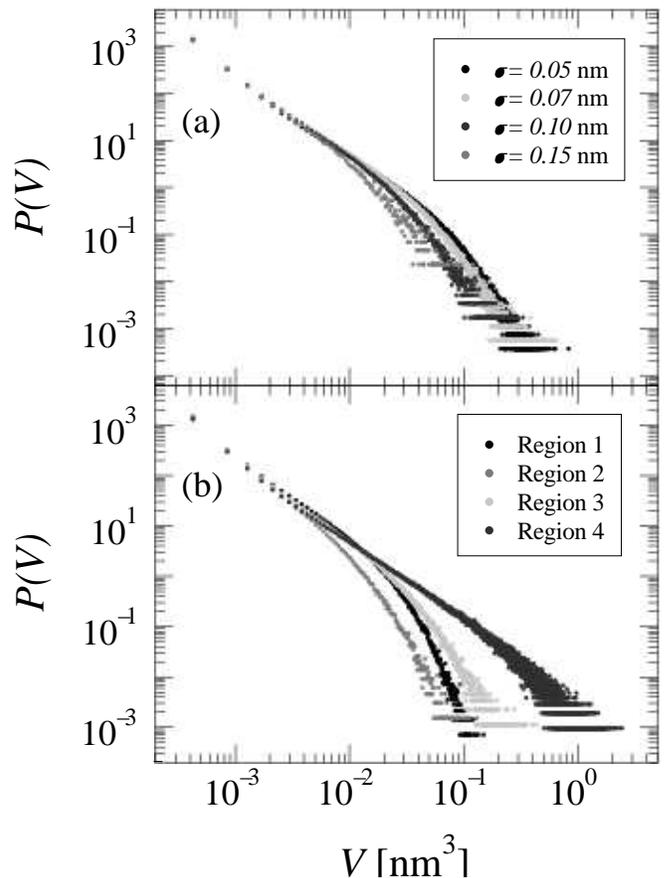}
\caption{\label{fig_num_regrad}Effect of penetrant size and
region on void size distribution.
(a)~Size distributions for different penetrant sizes for the case 
$\chi = 20.3$\,\% and Region 3.
(b)~Size distributions in different regions for $\chi = 20.3$\,\% and 
$r = 0.09$\,nm. The relative errors are similar to those in
Fig.~\ref{fig_num_cho}.}
\end{figure}

Having already mentioned the impact of penetrant size and region
on the void size distributions, let us discuss them at some more
length. Void size distributions at $\chi = 20.3$\,\% in Region 3 
for penetrants with radii ranging from $0.05$\,nm to $0.15$\,nm 
are displayed in Fig.~\ref{fig_num_regrad}(a).
We can immediately draw the obvious conclusion that large penetrants see
fewer and smaller voids than small ones. As Region 3 with 
$r \geq 0.05$\,nm is below any percolation threshold, 
all cases look very similar: power-law behavior with an exponent 
$1.5$ at small and intermediate $V$ and a cross-over to 
exponential behavior at larger $V$. The cross-over
moves to smaller $V$ with an increasing $r$, as we are 
getting further away from the xyz percolation threshold.

The effect of region on the case of $\chi = 20.3$\,\% and
$r = 0.09$\,nm is shown in Fig.~\ref{fig_num_regrad}(b).
The smallest number of voids, as well as the smallest voids,
can be found in the tightly packed headgroup region or Region 2.
As expected, Region 4 features much larger voids than any 
other region. All curves shown in this figure agree well
with the predictions of the bulk three-dimensional percolation theory
below the percolation threshold.

\subsection{Shape}

Using PCA we can extract $\sigma_1$, $\sigma_2$, and $\sigma_3$,
which are proportional to the lengths of the principal axes 
of an ellipsoidal void. From these we can compute 
$P(\sigma_1/\sigma_2,\sigma_2/\sigma_3)$,
a distribution describing the probability that a void has
given values of $\sigma_1/\sigma_2$ and $\sigma_2/\sigma_3$. The distribution 
has been normalized such that integration over it gives unity. As
visualizing and comparing different 
$P(\sigma_1/\sigma_2,\sigma_2/\sigma_3)$ can be difficult, we have also
computed $P(\sigma_i/\sigma_j)$ describing the probability of finding
a void with a given value of $\sigma_i / \sigma_j$ irrespective of
$\sigma_k$, $k \neq i,j$. These distributions have been scaled
such that integration over them gives unity.

Interpretation of the probability distributions will give us
information on the shape of the voids for which
$4 \times 10^{-3}$\,nm$^3$ $< V < 0.13$\,nm$^3$, 
see Sect.~\ref{sect_pca}. If $\sigma_1/\sigma_2 
\approx \sigma_2/\sigma_3 \approx 1$, the void is 
spherical. In case $\sigma_1 \gg \sigma_2 \approx \sigma_3$, we
are dealing with a roughly cigar-shaped void. Finally, if 
$\sigma_1 \approx \sigma_2 \gg \sigma_3$, our void is
essentially disk-like.

\begin{figure}[h]
\centering
\includegraphics[width=\columnwidth,clip=true]%
{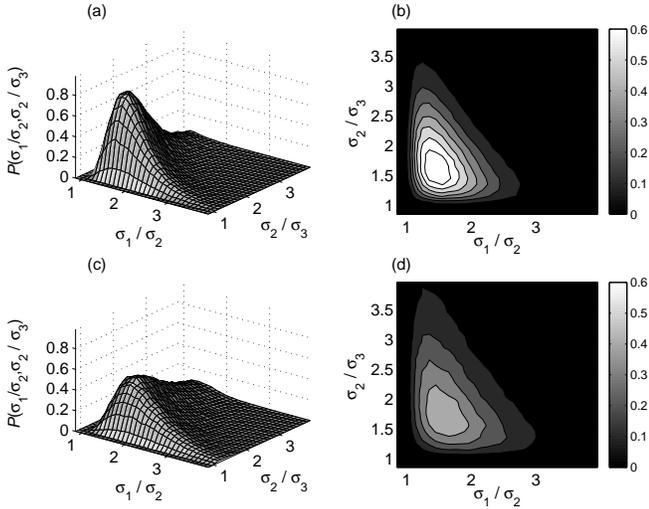}
\caption{\label{fig_rama_reg3_rad05}$P(\sigma_1/\sigma_2,\sigma_2/\sigma_3)$
at $\chi = 0$\,\% and $\chi = 29.7$\,\% in Region 3. The penetrant size
is $0.05$\,nm.
(a)~Surface plot at $\chi = 0.0$\,\%.
(b)~Contour plot at $\chi = 0.0$\,\%.
(c)~Surface plot at $\chi = 29.7$\,\%.
(d)~Contour plot at $\chi = 29.7$\,\%. The relative 
errors are smaller than ten percent.}
\end{figure}

The effect of cholesterol on $P(\sigma_1/\sigma_2,\sigma_2/\sigma_3)$ in 
Region 3 with $r = 0.05$\,nm is portrayed in 
Fig.~\ref{fig_rama_reg3_rad05}. Cholesterol does
not influence the shape of the voids in any other region, and other
penetrant sizes give very similar results (data not shown). 
Panels (a) and (b) show the probability distribution for pure DPPC
and panels (c) and (d) for a bilayer with $29.7$\,\% cholesterol.
In both cases elongated voids dominate the distributions:
spherical or nearly spherical voids are rare. The presence
of cholesterol makes the voids more elongated, i.\,e., larger values
of $\sigma_1/\sigma_2$ and $\sigma_2/\sigma_3$ occur with a higher
probability. 

\begin{figure}[h]
\centering
\includegraphics[width=\columnwidth,clip=true]%
{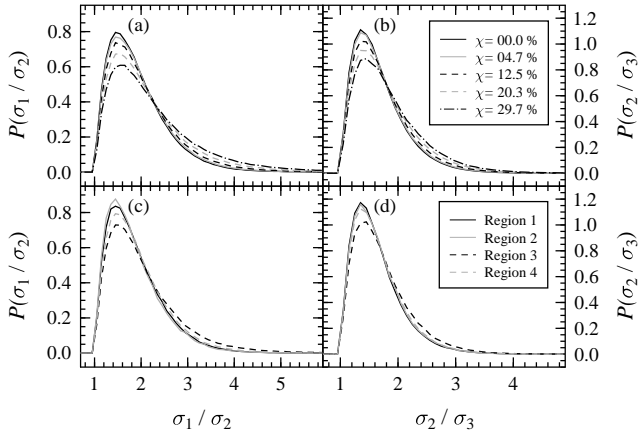}
\caption{\label{fig_shape}$P(\sigma_i/\sigma_j)$ at different cholesterol
concentrations and in different regions.
(a)~$P(\sigma_1/\sigma_2)$ in Region 3 for $r = 0.05$\,nm. 
(b)~$P(\sigma_2/\sigma_3)$ in Region 3 for $r = 0.05$\,nm. 
(c)~$P(\sigma_1/\sigma_2)$ for $\chi = 20.3$\,\% and $r = 0.09$\,nm.
(d)~$P(\sigma_2/\sigma_3)$ for $\chi = 20.3$\,\% and $r = 0.09$\,nm.
The relative errors are smaller than two per 
cent when $\sigma_i / \sigma_j < 3$, and
between two and five percent for $3 < \sigma_i / \sigma_j < 6$.}
\end{figure}

The elongating effect of cholesterol can be further studied by
focusing on $P(\sigma_1/\sigma_2)$ and $P(\sigma_2/\sigma_3)$
shown in panels (a) and (b) of Fig.~\ref{fig_shape}. These 
figures show a systematic average elongation with an increasing 
cholesterol content: the larger the concentration, the more
elongated voids we see.

Figure~\ref{fig_shape} also depicts the behavior of 
$P(\sigma_1/\sigma_2)$ and $P(\sigma_2/\sigma_3)$ in different 
regions, see panels (c) and (d). In this case
the bilayer contains $20.3$\,\% cholesterol and $r = 0.09$\,nm,
which represents well all penetrant sizes. Regions 1 and 2 are 
very similar to each other. Most voids larger 
than the lower cutoff $V = 4 \times 10^{-3}$\,nm$^3$ 
have been taken into account in PCA, since the vast majority 
of voids is smaller than the upper cutoff $V = 0.13$\,nm$^3$, 
see Fig.~\ref{fig_num_regrad}. Region 4 appears to differ from 
Regions 1 and 2 only slightly. This is true for small voids with 
$V < 0.13$\,nm$^3$. We should, however, keep in mind that Region 4 
also contains larger voids, see Fig.~\ref{fig_num_regrad}. The shapes 
of these holes are not ellipsoidal but more complex. 
The voids in Region 3 are more elongated than elsewhere. When the cholesterol
concentration increases, this effect becomes slightly more pronounced.
If $\chi = 0$\,\%, all regions look approximately similar.

\subsection{Orientation\label{sect_shape}}

The orientations of the principal axes of the ellipsoidal voids
can be extracted from PCA. We have measured the cosine of the
angle $\phi$ between the longest axis of an elongated ellipsoid and 
the bilayer normal. When the longest axis is oriented along the
bilayer normal $\phi = 0$. In principle, $\phi \in [0,\pi)$, but
due to the symmetry properties of ellipsoids the angles $\phi$
and $\pi - \phi$ are equivalent. We will therefore confine ourselves to 
$\phi \in [0,\pi/2]$. To describe the orientation of the voids,
we have computed the quantity $P(\cos \phi)$, the probability
that a given void is oriented such that the cosine of the
angle between its longest axis and the bilayer normal is $\cos \phi$.
The probability distribution has been normalized to give unity when 
integrated over, $\int_0^1 \mathrm{d}(\cos \phi) P(\cos \phi) = 1$. 
Should we wish to think in terms of $\phi$ rather than $\cos \phi$,
averages are computed by integrating over $\phi$ and the weight
of a given angle is $P(\cos \phi) \sin \phi$. 

\begin{figure}[h]
\centering
\includegraphics[width=\columnwidth,clip=true]%
{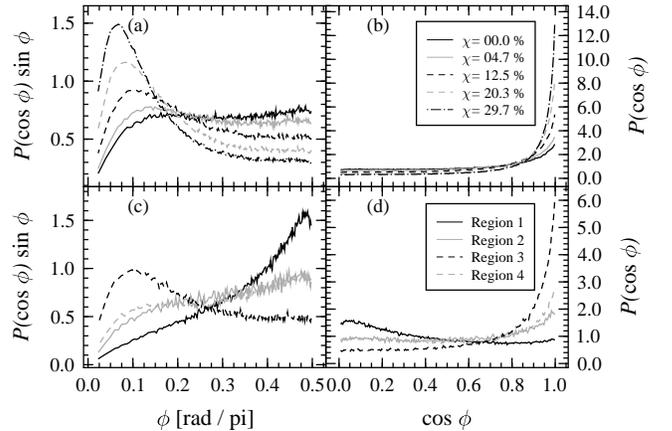}
\caption{\label{fig_tilt}Orientational distributions at different cholesterol
concentrations and in different regions.
(a)~$P(\cos \phi) \sin \phi$ in Region 3 for $r = 0.05$\,nm.
(b)~$P(\cos \phi)$ in Region 3 for $r = 0.05$\,nm .
(c)~$P(\cos \phi) \sin \phi$ for $\chi = 20.3$\,\% 
and $r = 0.09$\,nm.
(d)~$P(\cos \phi) \phi$ for $\chi = 20.3$\,\% and
$r = 0.09$\,nm. The relative 
errors are of the order of five percent for both
$P(\cos \phi)$ and $P(\cos \phi) \sin \phi$.}
\end{figure}

The impact of cholesterol concentration on $P(\cos \phi)$ and 
$P(\cos \phi) \sin \phi$ depends on which region of the bilayer 
we study. There are hardly any changes in either
Region 1 or 2. The effect of cholesterol is strongest in Region 3,
and largely independent of $r$. The case of $r = 0.05$\,nm
is portrayed in panels (a) and (b) of Fig.~\ref{fig_tilt}. In pure DPPC,
angles $\phi > 0.1 \pi$ are favored, while orientation along the bilayer 
normal hardly occurs at all. At $\chi = 29.7$\,\%, the angle with
the largest weight is $\phi = 0.06 \pi$. Orientation along the bilayer 
normal or close to it is favored, and perpendicular orientation
is rare. At cholesterol concentrations between 0 and 29.7\,\%,
the orientational distributions change systematically from
one extreme to the other. The presence of cholesterol can therefore
be said to promote orientation along the bilayer normal in Region 3.

In Region 4 the effects of cholesterol are less 
pronounced and depend on $r$ (data not shown). Between 
the xy and xyz percolation thresholds, e.\,g., in the 
case of $r = 0.05$\,nm, orientation along the
bilayer normal is favored. An increasing cholesterol concentration
seems to accentuate this effect. Below the xy percolation threshold, for
$r = 0.09$\,nm and larger, perpendicular orientation is 
favored and few voids are oriented along the bilayer normal,
see panel (c) of Fig.~\ref{fig_tilt}. Cholesterol has a minor effect here, 
suppressing the weight of intermediate angles $\phi \approx 0.25 \pi$
and making perpendicular orientational slightly more probable.
As discussed in the end of Sect.~\ref{sect_shape}, it should be kept
in mind that these distributions have been computed for voids smaller than
$V = 0.13$\,nm$^3$. For instance in the case of $r = 0.05$\,nm, Region 4
contains a large percolating cluster of a complex shape,
oriented more or less in the plane of the bilayer.

The situation in different regions at $\chi = 20.3$\,\% and for 
$r = 0.09$\,nm is portrayed in panels (c) and (d) of Fig.~\ref{fig_tilt}.
The effect of changing $\chi$ and $r$ has been described above.
In Regions 1 and 2, the orientational distribution of
voids is very broad with a peak close to $\phi = \pi/2$. 
Angles near $\phi \approx 0$ hardly occur at all. In Region 3, orientation 
along the bilayer normal or close to it is favored. 
Region 4, as described above, contains mostly voids that 
are oriented perpendicular to the bilayer normal.

\subsection{Size, Shape, and Orientation}

So far we have considered the shape and orientation of
all voids irrespective of their size. In this section we shall
investigate whether voids of different sizes differ in average 
shape or orientation. For this purpose we have computed
$P(V,\sigma_1/\sigma_2)$, the probability of finding a void of volume
$V$ with a given value of $\sigma_1 / \sigma_2$, and $P(V,\cos \phi)$,
the probability of finding a void of volume $V$ with a given $\cos \phi$.
Both probability distributions have been normalized such that
integration over them results in unity.

Unfortunately, $P(V,\sigma_1/\sigma_2)$ and $P(V,\cos \phi)$
are quite noisy and extremely difficult to visualize. Therefore we
have calculated the average values of $\sigma_1 / \sigma_2$ and
$\cos \phi$ for each $V$, defined as follows:
\begin{eqnarray} \label{eq_fixed_V}
\langle \sigma_1 / \sigma_2 \rangle_V & \equiv &
\frac{\int_1^\infty \mathrm{d} (\sigma_1/\sigma_2) P(V,\sigma_1/\sigma_2)
 \sigma_1/\sigma_2}{\int_1^\infty \mathrm{d} 
(\sigma_1/\sigma_2) P(V,\sigma_1/\sigma_2)}, \\ 
\langle \cos \phi \rangle_V & \equiv &
\frac{\int_0^1 \mathrm{d} (\cos \phi) P(V,\cos \phi) \cos \phi}
{\int_0^1 \mathrm{d} (\cos \phi) P(V,\cos \phi)}.
\end{eqnarray}
These might still be somewhat noisy, as the number of voids of a 
given size $V$ may be quite small.

In analyzing $\langle \sigma_1 / \sigma_2 \rangle_V$
and $\langle \cos \phi \rangle_V$, we shall focus on the effects of 
cholesterol in Region 3 and for $r = 0.05$\,nm. Based
on our studies of $P(\sigma_1/\sigma_2)$, $P(\cos \phi)$, and
related quantities, we expect the impact of cholesterol to
be most significant in this region. Further, we do not anticipate
our results to change significantly with $r$.

\begin{figure}[h]
\centering
\includegraphics[width=\columnwidth,clip=true]%
{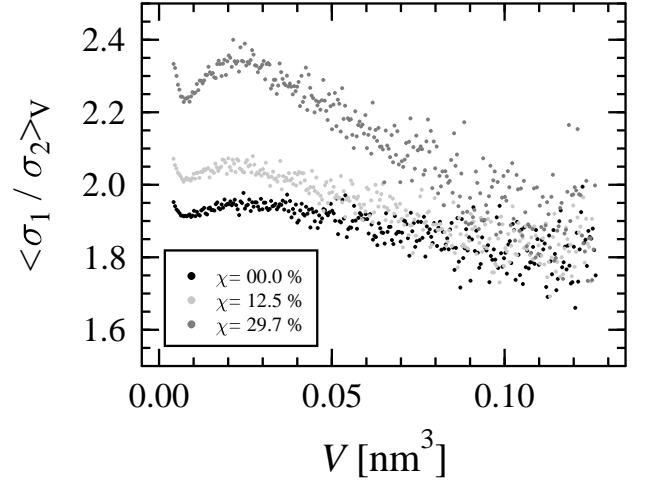}
\caption{\label{fig_shapevol}Averages $\langle \sigma_1 / \sigma_2 \rangle_V$ 
as functions of $V$ in Region 3 and for $r = 0.05$\,nm.
The relative errors grow with $V$: for $V < 0.05$\,nm$^3$ the errors 
are smaller than three percent, for 0.05\,nm$^3$ $< V < 0.10$\,nm$^3$ 
they grow linearly in the range $4-10$\,\%, and for larger $V$ they 
can be up to 20\,\%.}
\end{figure}

The behavior of $\langle \sigma_1 / \sigma_2 \rangle_V$ with $V$
is shown in Fig.~\ref{fig_shapevol}. In the case of pure DPPC,
our data indicate that voids of all sizes should
have approximately similar shapes. With $29.7$\,\% cholesterol 
small voids with $V < V_\mathrm{chol} / 10$ seem to be 
more elongated than larger ones. The shapes of large voids
are little affected by an increasing cholesterol concentration.
Based on these data, we can conclude that the presence 
and amount of cholesterol appear to mainly affect the shape of small voids. 

\begin{figure}[h]
\centering
\includegraphics[width=\columnwidth,clip=true]%
{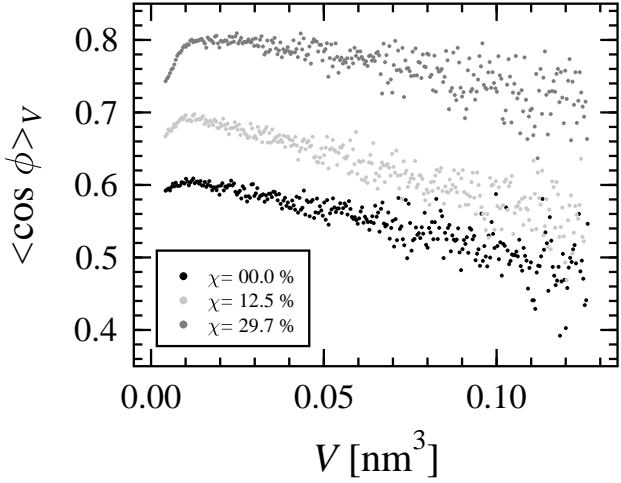}
\caption{\label{fig_cosvol}Averages $\langle \cos \phi \rangle_V$ 
as functions of $V$ in Region 3 and for $r = 0.05$\,nm.
The relative errors grow with $V$: for $V < 0.01$\,nm$^3$ the errors 
are smaller than a percent, for 0.01\,nm$^3$ $< V < 0.05$\,nm$^3$ 
smaller than five percent, for 0.01\,nm$^3$ $< V < 0.05$\,nm$^3$ 
smaller than ten percent, and for larger $V$ up to 25\,\%.}
\end{figure}

Figure~\ref{fig_cosvol} depicts the effect of cholesterol on
$\langle \cos \phi \rangle_V$ vs.\,$V$. At all $\chi$, 
the voids become on the average slightly more 
perpendicularly oriented with an increasing $V$. An increasing cholesterol
concentration makes voids of all sizes adopt average orientations closer 
to the direction of the bilayer normal. We can therefore conclude
that an increasing cholesterol concentration affects 
the orientation of at least the voids 
with $4 \times 10^{-3}$\,nm$^3$ $< V < 0.13$\,nm$^3$.

\begin{figure}[h]
\centering
\includegraphics[width=.7\columnwidth,clip=true]%
{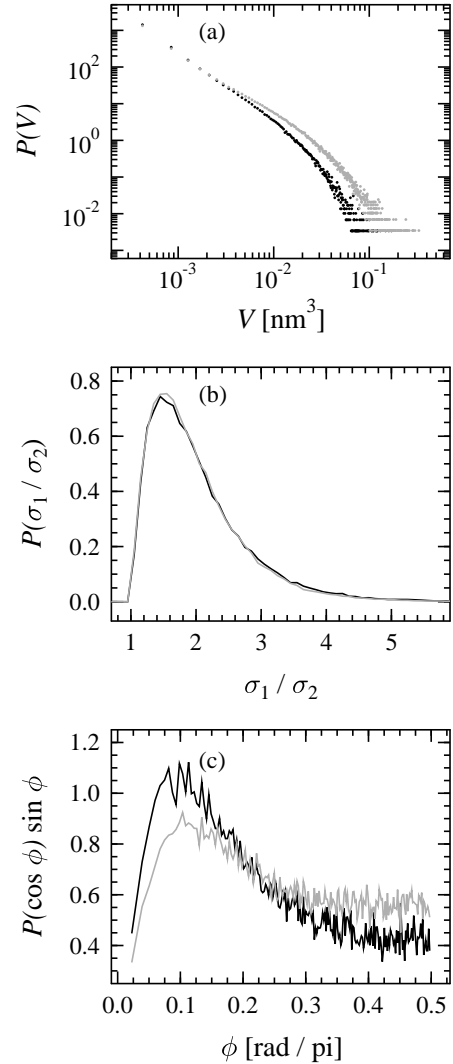}
\caption{\label{fig_local}Effect of proximity of cholesterol
molecules at $\chi = 20.3$\,\% with $r = 0.10$\,nm. 
The data for voids close to cholesterols are shown in black 
and those for voids far from cholesterols in gray.
(a)~Normalized void size distributions.
(c)~$P(\sigma_1 / \sigma_2)$.
(b)~Orientational distributions.}
\end{figure}

\subsection{Local Effects}

We have seen that a finite cholesterol concentration changes 
the properties of voids in a bilayer. It is natural to ask
whether or not the changes are more pronounced in the close 
vicinity of cholesterol molecules. In principle, investigating 
the locality of the changes cholesterol brings about is simple: 
we just have to single out the voids whose CMs are 
located near the CM of a cholesterol molecule, and 
compare their properties to voids whose CMs are nowhere near 
cholesterols. There are, however, a few catches that 
should be avoided.

First, studying arbitrarily large voids is meaningless. The CM of a very 
large void may or may not be close to the CM of a cholesterol,
even though some parts of the void might be in the vicinity of one
or several cholesterols. In any event, such large clusters cannot
very well be said to be localized. Hence, we have confined ourselves
to voids that are smaller than a cholesterol molecule, i.\,e.,
voids with $V < V_\mathrm{chol}$.

An additional problem is choosing a suitable criterion for
the proximity of a void to a cholesterol molecule. The criterion 
should not be too strict, leading to automatic exclusion of all 
large and perpendicularly oriented voids from the class of voids close 
to cholesterol. This can be ascertained by monitoring the principal radii
of voids and choosing the criterion based on these. The choice can
be facilitated by restricting the study to Region 3, where the voids 
will, by definition, be in the same region as the steroid ring structures. 
This allows us to consider distances that are projected to 
the plane of the bilayer.

We have chosen to focus on concentration $\chi = 20.3$\,\%, 
and the penetrant radius has been set to $r = 0.10$\,nm. 
After investigating the principal axes of voids in this case, 
we have decided to regard voids with CMs within $0.8$\,nm of a 
CM of a cholesterol molecule, making sure that both are 
in the same monolayer, as proximate to cholesterol.
This decision is based on two observations: the radius
corresponding to the thickest part of a cholesterol molecule
in our study is approximately $0.3$\,nm,\cite{Fal04} and the 
vast majority of voids in our current setup have principal radii 
smaller than $0.5$\,nm.

The results are shown in Fig.~\ref{fig_local}. Panel (a)
features the size distributions computed for voids in the vicinity 
of cholesterols and for those far from cholesterols. To facilitate 
comparison, these distributions have been normalized by 
the average number of voids close to cholesterols in 
Region 3 and by the average number of voids far from cholesterol in 
Region 3. The resulting normalized distribution $P(V)$ is 
such that it gives unity when integrated over. Voids with radii 
smaller than $0.1$\,nm are unaffected by the 
proximity of cholesterol. Larger voids, however, are affected:
the vicinity of cholesterol reduces the probability of finding
a void with $V \approx V_\mathrm{chol} / 10$ by a little less than a 
factor of ten. Further, near cholesterols there are no voids
with $0.1$\,nm$^3$ $< V < V_\mathrm{chol}$.

Is the size distribution for voids far from cholesterol 
coincident with the size distribution for Region 3 of a 
pure DPPC bilayer? Considering voids with $V < V_\mathrm{chol}$ 
and normalizing by the total number of voids in 
each case, we obtain distributions that can be compared 
directly. It turns out that the two are nearly identical. 
Hence, further away from cholesterols the size distribution
of voids is unaffected by the presence of cholesterol and practically
identical to the corresponding distribution in pure DPPC.

The effect of the proximity of cholesterol molecules on void
shape is portrayed in panel (b) of Fig.~\ref{fig_local}. It is 
evident that the shape of voids does not depend on the vicinity of 
cholesterols. Orientation of voids near cholesterols and far from them 
is portrayed in panel (c) of the same figure. The voids that
are located close to cholesterols tend to be more oriented along the 
bilayer normal, while the ones far from cholesterols show
a broader orientational distribution.

\section{Concluding Discussion}

We have studied the impact of cholesterol on the voids in
simulated bilayers consisting of DPPC. Using data from
atomistic molecular dynamics simulations, we have investigated the behavior of
the total free volume with an increasing cholesterol concentration, 
extracted the void size distributions in different parts of the bilayer, 
and considered the effect of cholesterol on the shape and orientation of the 
voids. We have also probed whether or not the effects of 
cholesterol on the properties of voids are localized to the 
near vicinity of cholesterol molecules. In interpreting the data, we
have, in the spirit of Ref.~\onlinecite{Mar96a}, made
use of three-dimensional isotropic percolation theory.

Our findings support the reduction of the total free volume
with an increasing cholesterol content. The effects of cholesterol
on individual voids appear to be the most prominent in those
parts of the bilayer where cholesterol's steroid ring structures
and the DPPC tails are located. Here cholesterol reduces the 
connectivity of free volume, thus altering the void size distributions. 
These changes are perceptible already at low cholesterol concentrations 
and become pronounced with approximately 20\,\% cholesterol in the bilayer.
For instance a concentration of 29.7\,\% leads to complete 
removal of the voids that are approximately of the same size as 
a cholesterol molecule. An increasing cholesterol concentration 
also makes the voids in the region containing the steroid rings more 
elongated, and transforms the isotropic orientational 
distribution of voids present in pure DPPC to an anisotropic 
one favoring orientation of voids along the bilayer normal. All these 
effects appear to be localized to the proximity of cholesterol 
molecules, rather than being uniformly distributed
to the whole bilayer.

Reasons for the changes in the region with the cholesterol ring
structures and the hydrocarbon tails of DPPCs can be found in
previous computational studies on DPPC\,/\,cholesterol 
systems.\cite{Chi02,Fal04,Hof03,Rog01a,Smo99} Cholesterol 
increases the order parameters of the
hydrocarbon tails by simultaneously reducing their tilt and the number
of gauche-defects present in them. The straightening of the tails should
affect both the void size distributions and, in particular, the
shape and orientation of the voids. This explanation is supported by
the fact that the effects of cholesterol are confined to the proximity
of cholesterol molecules: the effect of cholesterol on the ordering of
the phospholipid tails is much stronger in the vicinity
of cholesterol molecules.\cite{Jed03a} In addition to the 
ordering, there is another, more obvious factor, which should certainly
influence the void size distributions. The bulky cholesterol molecule 
simply fills empty spaces of its own size in the bilayer, as suggested by
Almeida et al.\cite{Alm92}

Our results for pure DPPC are in good qualitative agreement with 
the findings of Marrink et al.,\cite{Mar96a} although their 
study has been performed at $T = 350$\,K and is based on 
80\,ps of data. The form of our void 
size distributions in different regions and for different 
permeants is in surprisingly good agreement with theirs. Their  
measure for the shape of the voids is different from ours,
not enabling them to fully distinguish between elongated and fractal voids. 
In addition, they study voids of all sizes, while we have
restricted our studies on the shape and orientation of
voids to the range $4 \times 10^{-3}$\,nm$^3$ $< V < 0.13$\,nm$^3$.
Their qualitative conclusion is that larger voids may be more
elongated and\,/\,or more fractal than small ones. We, on the other hand, 
find that the shape of the voids depends only weakly on the size, and that
the voids become slightly less elongated with an increasing volume. 
As for orientation, despite the different measures, their conclusions are in
accord with ours. They observe that the orientational distribution of voids
is most anisotropic is in Region 3, while Region 1 features
an almost completely isotropic distribution.

We also agree with Marrink et al.\cite{Mar96a} in that the complexity 
of the properties of the voids in a phospholipid bilayer
imply that simple free area theories cannot be used for quantitative 
predictions. However, free volume arguments may be used for
qualitatively explaining diffusion and permeation in bilayers.
An increasing cholesterol concentration in DPPC bilayers
leads to monotonously decreasing lateral diffusion coefficients for both 
DPPCs and cholesterols.\cite{Fal04} Based on our current findings
on the properties of voids, this is to be expected. Intuitively,
smaller and fewer voids, which are oriented along the bilayer normal,
should slow down lateral diffusion of lipids and sterols. The details
of what role the voids play for the various diffusion and permeation 
mechanisms in bilayers are challenging questions that 
large-scale simulation studies may elucidate in the near future.

In addition to the static properties of voids considered in this study,
also the dynamic properties should be important to dynamic processes
in bilayers. It would be interesting to study the dynamics of void 
formation and the movement of individual voids. Such studies should
give us further insight into mechanisms of lateral diffusion and solute 
permeation in membranes. 

\begin{acknowledgments}
We are grateful for Teemu Murtola and Mikko Alava
for fruitful discussions. This work has, in part, been 
supported by the Academy of Finland 
through its Center of Excellence Program (E.\,F., I.\,V.), the 
National Graduate School in Materials Physics (E.\,F.), 
the Academy of Finland Grant Nos.~54113, 00119 (M.\,K.), 80246 (I.\,V.),
and 80851 (M.\,H.), and the Jenny and Antti Wihuri Foundation (M.\,H.).
M.\,P. would like to acknowledge the support through
the Marie Curie fellowship No.~HPMF--CT--2002--01794. 
We would also like to thank the Finnish IT Center for Science 
and the HorseShoe (DCSC) supercluster computing facility 
at the University of Southern Denmark for computer resources. 
\end{acknowledgments}


\end{document}